# Experimental Realization of Acoustic Bianisotropic Gratings

Steven R. Craig[1,†], Xiaoshi Su[2,†], Andrew Norris[2], and Chengzhi Shi[1,3,*]

1 Meta Acoustics Lab, GWW School of Mechanical Engineering, Georgia Institute of Technology, Atlanta, GA 30332, USA
2 Department of Mechanical and Aerospace Engineering, Rutgers University, Piscataway, NJ 08854, USA
3 Parker H. Petit Institute for Bioengineering and Biosciences, Georgia Institute of Technology, Atlanta, GA, 30332, USA

Acoustic bianisotropic materials couple pressure and local particle velocity fields to simultaneously excite monopole and dipole scattering, which results in asymmetric wave transmission and reflection of airborne sound. In this work, we systematically realize an arbitrarily given bianisotropic coupling between the pressure and velocity fields for asymmetric wave propagation by an acoustic grating with inversion symmetry breaking. This acoustic bianisotropic grating is designed by optimizing the unit cells with a finite element method to achieve the desired scattering wavevectors determined by the bianisotropic induced asymmetric wave propagation. The symmetry and Bloch wavevectors in the reciprocal space resulted from the grating are analyzed, which match with the desired scattering wavevectors. The designed structures are fabricated for the experimental demonstration of the bianisotropic properties. The measured results match with the desired asymmetric wave scattering fields.

Manipulating acoustic wave propagation facilitates the development of novel applications in numerous fields, including lensing [1], noise control [2], and high-intensity focused ultrasound therapies [3, 4]. Due to the reversible nature of wave functions, classical acoustic waves propagate with symmetries in both time and space. Asymmetric transmission and reflection stem from breaking these symmetries for the realization of nonreciprocal acoustic propagation, inversion symmetry violation, or parity-time symmetry. A common way to achieve nonreciprocity in acoustics is by altering the frequency of incident waves using nonlinear materials and filtering unwanted frequencies with sonic crystals, enabling one-way acoustic propagation [5]. Using nonlinearities to achieve asymmetric wave propagation has also been demonstrated with active elements by coupling Helmholtz resonators with a piezoelectric material and a nonlinear circuit [6]. Alternative techniques to realize nonreciprocity have been accomplished with circular flow in a resonant ring cavity to introduce a momentum bias on the propagating wave [7]. This is an acoustic analog to the Zeeman Effect in electromagnetics, achieved by a magnetic field. Appling this approach of nonreciprocal circular flow to an array of cylindrical sonic crystals leads to the creation of a topologically protected edge state with one-way acoustic wave propagation, which is an acoustic analogy of the quantum Hall effect [8-10]. Besides breaking the time reversal symmetry with circular flow or synthetic angular momentum, topologically protected one-way edge states were also realized by breaking the inversion symmetry of phononic crystals that induces acoustic pseudospin, analogous to quantum spin Hall effect [11-13]. Generalizing topological acoustics in higher dimensions leads to the exploration of acoustic Weyl points and Fermi arcs, resulting in to asymmetric wave propagation in 3D [14-16]. In addition, asymmetric wave transmission has been observed in non-Hermitian structures with pure loss effects [17]. Non-Hermitian acoustic structures with parity-time symmetry, where the loss and gain materials were exactly balanced, was demonstrated to exhibit asymmetric wave reflection while maintaining total transmission on both sides of the structure at their exceptional points [18-20].

Bianisotropic materials enable asymmetric wave scattering with unitary efficiency in the bulk state compared to the existing approaches [21]. Bianisotropic properties were observed in electromagnetism where a coupling tensor that relates electric and magnetic fields with monopolar and dipolar moments in the scattered waves [22]. Acoustic bianisotropic materials couple the pressure and local particle velocity fields with monopole and dipole scattering resulting in asymmetric wave propagations [23-30]. The bianisotropic material was used as an effective sound barrier [31]. The coupling between the pressure and local particle velocity fields along with the scattered acoustic monopole and dipole is characterized by a polarizability tensor. In a 2D scenario, the polarizability tensor is a second order tensor given by

$$\begin{pmatrix} M \\ D_x \\ D_y \end{pmatrix} = \begin{pmatrix} \hat{\alpha}^p & \hat{\alpha}^{pv}_x & \hat{\alpha}^{pv}_y \\ \hat{\alpha}^{vp}_x & \hat{\alpha}^{vv}_{xx} & \hat{\alpha}^{vv}_{xy} \\ \hat{\alpha}^{vp}_y & \hat{\alpha}^{vv}_{yx} & \hat{\alpha}^{vv}_{yy} \end{pmatrix} \begin{pmatrix} p \\ v_x \\ v_y \end{pmatrix}, \quad (1)$$

where $M$ is the monopole scattering, $D_x$ and $D_y$ are the dipole scattering along the $x$ and $y$ axis, $p$ is the pressure field, $v_x$ and $v_y$ are the local particle velocity fields in the $x$ and $y$ directions, and $\hat{\alpha}^i_j$ are the elements of the polarizability tensor with $i$ denoting the coupling between the pressure and local particle velocity fields and *j* denoting the coupling axes, respectively. Given an arbitrary polarizability tensor, one can obtain the scattering field of a bianisotropic particle [32]. For example, the scattering of a bianisotropic particle allowing for both forward and backward scattering, whose polarizability tensor is a function of frequency given by Fig. 1a, is obtained in Fig. 1b. Similarly, the scattering of another bianisotropic particle with a polarizability tensor given by Fig. 1c only permitting backwards scattering is shown in Fig. 1d.

In this work, we aim to realize an acoustic bianisotropic grating with its polarizability tensor given by Fig. 1a at 6 kHz (wavelength equal to 5.7 cm). Because the bianisotropic grating we intend to fabricate is linear, the wavelength and frequency remain unchanged during the scattering process, and the incident and scattered wavevectors retain on a circle with its radius given by $k = 2\pi/\lambda = 109.9$ m$^{-1}$ (Figs. 2a and 2b). The Bloch wavevectors of the grating required to achieve this asymmetric wave propagation in Fig. 1b is given by the difference between the incident and scattered wavevectors shown by the black arrows in Figs. 2a and 2b. A -45-degree incident wave is completely reflected by the bianisotropic grating back towards the source that requires a Bloch wavevector $(-\sqrt{2}k, \sqrt{2}k)$ given by the structure (Fig. 2a). The time reversal of the scattering process for this incidence is identical to itself (Fig. 2a). On the other hand, a 45-degree incident wave is refracted with a total transmission in the -45-degree direction, which requires a Bloch wavevector $(0, -\sqrt{2}k)$ from the grating shown in Fig. 2b. The time reversal of this incident case exhibits the same refraction, confirming the preservation of time reversal symmetry in the bianisotropic grating. The asymmetric wave scattering of the -45- and 45-degree incidents reveal the violation of parity symmetry in the *y* direction, indicating the inversion symmetry breaking of the bianisotropic grating about the *x*-axis. Thus, the mirrored Bloch wavevectors $(\sqrt{2}k, \sqrt{2}k)$ and $(0, \sqrt{2}k)$ about the x-axis are not provided by the grating. To realize this bianisotropic grating, we determine the grating periodicity from the scattered wave directions from the aforementioned polarizability tensor. The geometry of the unit cells are optimized based on a finite element method, which maximizes the scattering efficiency of the grating (Supplementary Information). The resulting bianisotropic grating consists of a periodic array of scatterers along the y-axis with 40.4 mm spacing. Each scatterer consists of a rectangle having a base of 39.5 mm and height of 14.8 mm adjacent to a second rectangle with a 19.5 mm base and 21.2 mm height centered above the first (Fig. 2c). In order to confirm that the grating elements provide the desired scattering in Fig. 1b, Fourier analysis of the symmetry properties of the scattered wavevectors in the reciprocal space is performed (Fig. 2d), which results in the wavevectors required by the scattering in Figs. 2a and 2b.

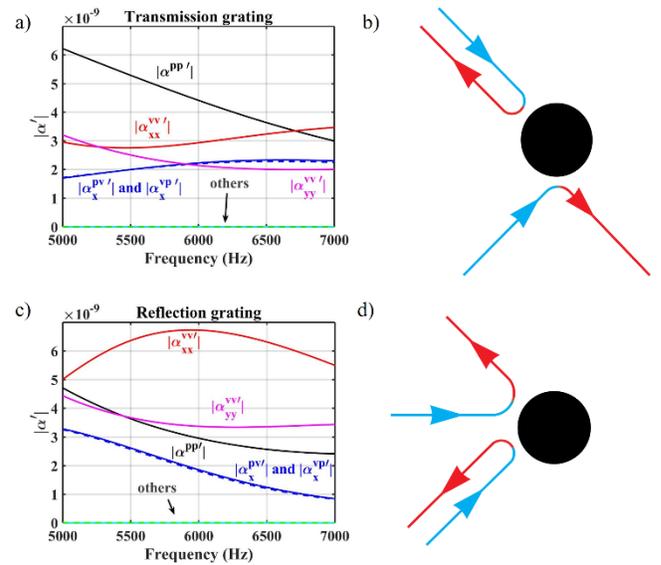

Figure 1: (a) The elements of the polarizability tensor of a bianisotropic particle that allows both forward and backward scattering as functions of frequency. (b) Asymmetric wave scattering for different incidences from the bianisotropic particle in (a). (c) The elements of the polarizability tensor of a bianisotropic particle with only backward scattering. (d) Asymmetric wave scattering for different incidences from the bianisotropic particle in (c). For both (b) and (d), the blue arrows represent incident and red arrows represent scattered waves, respectively.

To verify the bianisotropic properties of the designed grating experimentally, 25 aluminum grating elements were machined by CNC milling (Fig. 2e). In our experiment, the grating elements were arranged in a two-dimensional waveguide with a height of 2 cm bounded by two 1000 mm by 1500 mm acrylic sheets. Plane waves were generated by a speaker array consisting of twelve 17 mm diameter speakers spaced $\lambda/2$ apart, positioned at a 45-degree angle relative to the grating with the closest edge of the array being 500 mm away from the grating (Fig. 2e). In order to create a consistent signal, the pressure amplitude of the speakers was controlled by a digital, multichannel recorder while an omnidirectional microphone attached to a motorized positioner measured the pressure fields for both -45- and 45-degree incidence cases. The motorized positioner was controlled by a MATLAB script to scan two 450 mm by 800 mm areas, which are 10 mm away from the grating with a scan resolution of $\lambda/10$. A lock-in amplifier records the amplitude and phase of the acoustic waves in the scan area (Fig. 2e). To distinguish the incident and scattered waves, spatial 2D Fourier analysis is performed on the measured pressure fields, differentiating the waves with positive $k_x$ (Figs. 2f and 2h) and negative $k_x$ components (Figs. 2g and 2i). For -45-degree incidence, the acoustic wave is reflected back towards the source by the bianisotropic grating with pressure reflection coefficient $R = 0.72$ and no transmission through the grating (Fig. 2f and 2g). This experimental result matches the desired scattering of the bianisotropic structure in Fig. 1b with its polarizability tensor given by Fig. 1a. On the other hand, a 45-degree incidence results in a transmission at a -45-degree angle with pressure transmission coefficient $T = 0.78$ and no reflection from the grating (Figs. 2h and 2i), which matches the scattering of the bianisotropic material in Fig. 1b. The full wave simulations show near perfect reflection and transmission for the two incident cases, respectively (Supplementary Information), which indicates near perfect scattering efficiencies. The experimental error arises from the speaker array being an imperfect source that excites unwanted wavevectors. In addition, the finite dimensions of the waveguide possess open boundaries at the edges of the acrylic sheets resulting in unwanted reflections (as shown by the Fourier analysis in Supplementary Information). Further error arises from the fabrication and imperfect alignment of the grating. These experimental results confirm the feasibility for the realization of bianisotropic properties by breaking the inversion symmetry of the structure.

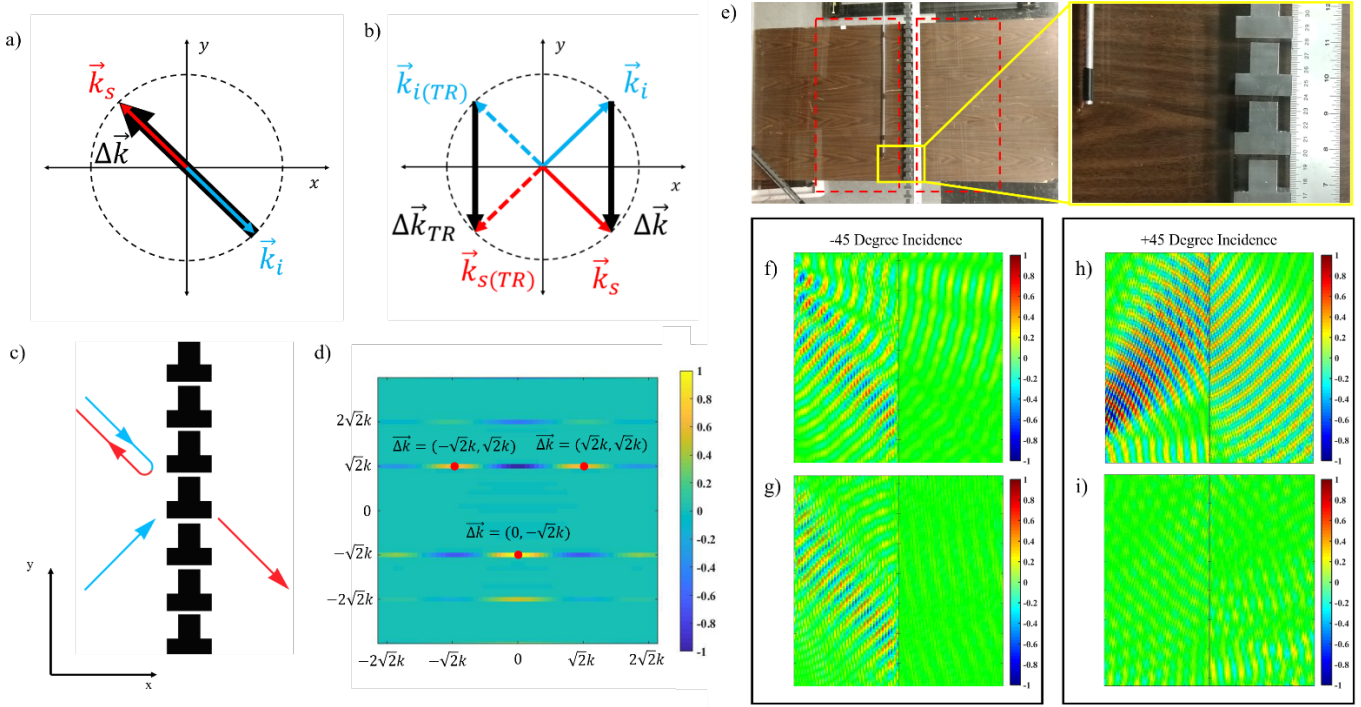

Figure 2: (a) Wave-grating interaction for the transmission case for -45-degree incidence in the reciprocal space. (b) Wave-grating interaction for +45-degree incidence as well as the time reversed case represented by the dashed lines. For both (a) and (b), the incident, scattered, and Bloch wavevectors are represented by $\vec{k}_i$, $\vec{k}_s$ and $\Delta k$, respectively. (c) Bianisotropic grating structure with asymmetric wave scattering in (a) and (b) determined by inverse Fourier analysis of the Bloch wavevectors. (d) 2D spatial Fourier transform of the grating geometry with the expected scattering wavevectors having the highest intensity. (e) Experimental set up for the bianisotropic transmission grating. The zoom out shows the detailed structure of the grating. The outlined rectangles with red dashed lines represent the scanned areas. (f) Measured pressure field of acoustic waves propagating with positive $k_x$ component for -45-degree incidence. (g) Measured pressure field with negative $k_x$ component for -45-degree. (h) Measured pressure field of acoustic waves propagating with positive $k_x$ component for +45-degree incidence. (i) Measured pressure field with negative $k_x$ component for +45-degree incidence. All the pressure fields in (f), (g), (h), and (i) are normalized by their corresponding maximum amplitude in each measurement.

The realization of a bianisotropic grating with polarizability defined by Fig. 1d at 6kHz is similar to the procedure previously executed. By analyzing the wave-grating interaction in reciprocal space, the Bloch wavevectors can be determined for each incidence case. We continue to use a bianisotropic grating with linear elements, the incident and scattered wavevectors are known to remain on a circle with radius $k = 109.9$ m$^{-1}$ (Fig. 3a and 3b). A wave with 45-degree incidence interacting with the bianisotropic grating results in a total reflection back towards the source requiring a Bloch wavevector of $(-\sqrt{2}k, -\sqrt{2}k)$ (Fig. 3a). This Bloch wavevector is identical to that of the time reversal propagation. For normal incidence, the acoustic wave is fully reflected towards 135-degree direction, which requires a Bloch wavevector of $(-\left(1+\frac{\sqrt{2}}{2}\right)k, \frac{\sqrt{2}}{2}k)$ (Fig. 3b). The time reversal case for this incidence requires an identical Bloch wavevector, confirming the time reversal symmetry for both incidence cases. The asymmetric wave propagation for different incidence reveals the violation of parity symmetry about the x-axis. Similar to the bianisotropic grating above, the scattered wave directions determine the grating periodicity while a finite element method is used to optimize the unit cell geometry and maximize the scattering efficiency (Supplementary Information). The resulting bianisotropic grating consists of an array of scatterers having an 80.8 mm periodicity. Each scatterer consists of two rectangles: the first having a base of 18.5 mm and height of 26.5 mm directly underneath the second rectangle with a base of 9.4 mm and height of 21.2 mm (Fig. 3c). Analyzing the grating geometries with spatial Fourier transform reveals the resulted Bloch wavevectors in Fig. 3d matching the scattering wavevectors from the symmetry analysis in Figs. 3a and 3b.

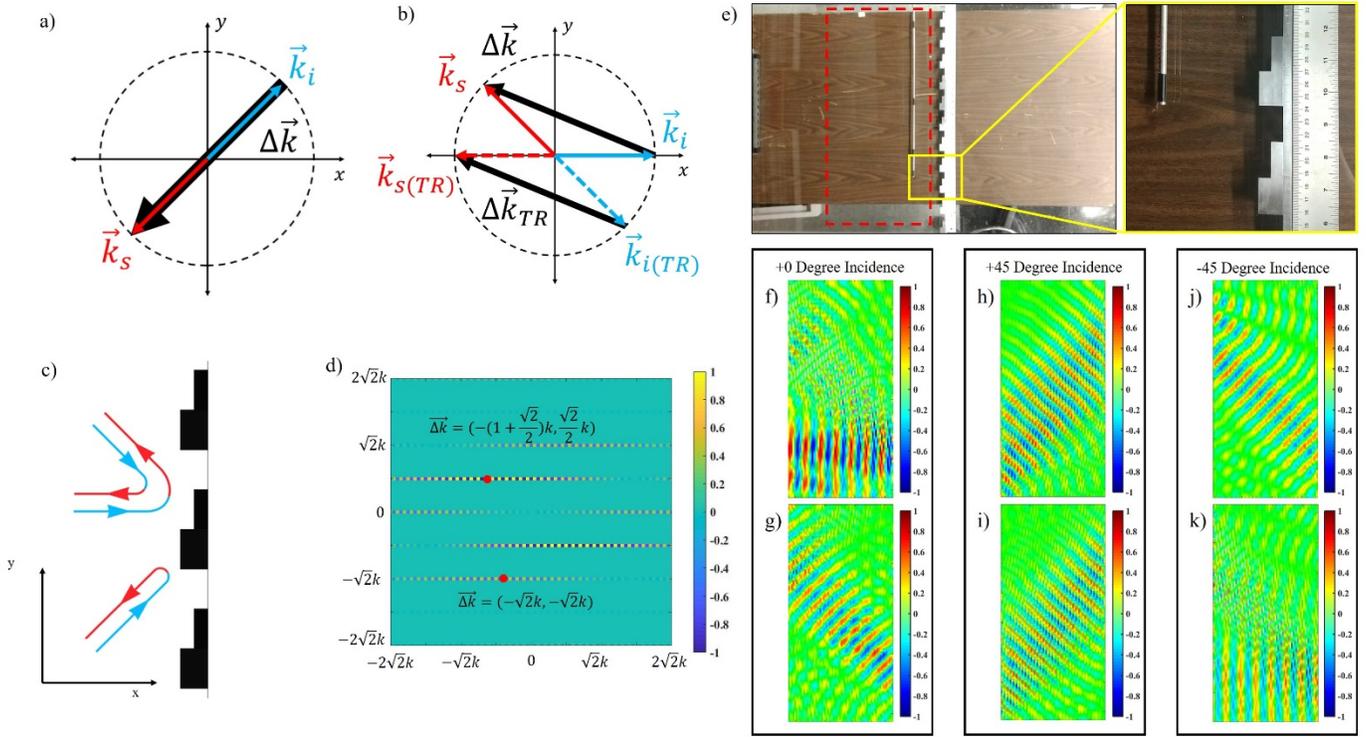

Figure 3: (a) Wave-grating interaction for the reflection case for 45-degree incidence in the reciprocal space. (b) Wave-grating interaction for normal incidence as well as the time reversed case represented by the dashed lines. For both (a) and (b), the incident, scattered, and Bloch wavevectors are represented by $\vec{k}_i$, $\vec{k}_s$ and $\Delta k$, respectively. (c) Bianisotropic grating structure with asymmetric wave scattering in (a) and (b) determined by inverse Fourier analysis of the Bloch wavevectors. (d) 2D spatial Fourier transform of the grating geometry with the expected scattering wavevectors having the highest intensity to the left of the inversion symmetry about the y-axis (e) Experimental set up for the bianisotropic reflection grating. The zoomed in image shows the detailed structure of the grating. The outlined rectangle with red dashed lines represents the scanned areas. (f) Measured pressure field of acoustic waves propagating with a positive $k_x$ component for normal incidence. (g) Measured pressure field with negative $k_x$ component for normal incidence. (h) Measured pressure field of acoustic waves propagating with positive $k_x$ component for +45-degree incidence. (i) Measured pressure field with negative $k_x$ component for +45-degree incidence. (j) Measured pressure field of acoustic waves propagating with positive $k_x$ component for -45-degree incidence. (k) Measured pressure field with negative $k_x$ component for -45-degree incidence. All the pressure fields in (f), (g), (h), (i), (j), and (k) are normalized by their corresponding maximum amplitude in each measurement.

The designed bianisotropic grating with thirteen elements are fabricated for the verification of the asymmetric wave scattering for the reflection only case (Fig. 3e). The grating elements were arranged linearly in the same waveguide described above. The speaker array was positioned in the same location for angled incidence, but was relocated to the end of the waveguide, 660 mm away from the grating for the normal incidence case. The motorized positioner scanned an identical rectangular area to measure the total pressure field with both incident and scattered waves. The 2D spatial Fourier analysis mentioned above is used to separate the incident and scattered waves. Experimental measurement shows a normal incidence interacting with the bianisotropic grating results in a reflection at a 135-degree angle with reflection coefficient $R = 0.76$ (Figs. 3f and 3g). The time reversal of this incidence is illustrated in Figs. 3j and 3k with a reflection coefficient $R = 0.88$. The 45-degree incident wave is reflected with a reflection coefficient $R = 0.84$ back towards the source (Figs. 3h and 3i). These experimental results match with the desired scattering of the bianisotropic material in Figs. 1e and 1f, whose polarizability tensor is given by Fig. 1d. Once again, the simulations show a near perfect reflection for each incident case (Supplementary Information). The unwanted wavevectors produced by the imperfect source array as well as the open boundaries at the edges of the waveguide affect the reflections in the experiment as shown in the Fourier analysis of the propagating waves in Supplementary Information. The fabrication and alignment errors of the grating also contribute to the experimental error.

In conclusion, we demonstrate a systematic approach to realize an arbitrarily given bianisotropic polarizability by a designed acoustic grating (for other arbitrary angles of incidence, see details in Supplementary Information). This is achieved by optimizing the geometry of the grating element to realize maximum scattering efficiency and confirming the asymmetric wave scattering of the desired bianisotropic properties in the reciprocal space through spatial Fourier analysis. The resulted parity symmetry violation of the acoustic wave propagation indicates the necessity of inversion symmetry breaking for the realization of the bianisotropic properties. The experimentally measured acoustic wave interactions with the designed bianisotropic grating match the desired scattering patterns of the given polarizability tensor. Our method paves the road for systematic realization of bianisotropic properties and asymmetric wave propagation with unitary efficiency, which is important for one-way acoustic filtering, sensing, and lensing.

[†] These authors contributed equally to this work
[*] Corresponding author.
chengzhi.shi@me.gatech.edu